# Magnetism and electronic dynamics in $CuCr_{2-x}Sn_xS_4$ spinels studied by transferred hyperfine fields at $^{119}Sn$ and muon spin rotation and relaxation

Elaheh Sadrollahi[1], Cynthia P. C. Medrano[2], Magno A.V. Heringer[2], E. M. Baggio Saitovitch[2], Lilian Prodan[3,4], Vladimir Tsurkan[3,4], and F. Jochen Litterst[5]

[1]Institut für Festkörper- und Materialphysik, Technische Universität Dresden, 01069 Dresden, Germany
[2] Centro Brasileiro de Pesquisas Físicas, Rio de Janeiro, 22290-180, Brazil
[3]Experimental Physics V, Center for Electronic Correlations and Magnetism, Institute of Physics, University of Augsburg, 86135 Augsburg, Germany
[4]Institute of Applied Physics, Moldova State University, MD 2028, Chisinau, Republic of Moldova
[5]Institut für Physik der kondensierten Materie, Technische Universität Braunschweig, 38106 Braunschweig, Germany

**Abstract**

We investigated magnetization, muon spin rotation and relaxation (µSR), and $^{119}Sn$ Mössbauer spectroscopy on Sn substituted $CuCr_{2-x}Sn_xS_4$ (x=0.03 and 0.08) spinel compounds. The magnetization and µSR results reveal similar additional low-temperature magnetic transitions around 80 K and 40 K as found for the undoped material, indicating a magnetic ground state deviating from a simple collinear ferromagnet. The observed changes in the Mössbauer spectra are less pronounced and are discussed in view of the different positions of the local probes $\mu^+$ and $^{119}Sn$ and their different magnetic coupling to the magnetic Cr lattice. Above 80 K, both µSR and Mössbauer spectra show temperature-dependent inhomogeneous broadening either due to structural or charge disorder and changing spin dynamics that can be related to a precursor magnetic phase above the well-defined static low-temperature phase.

## 1. Introduction

Our recent µSR studies and magnetization measurements on cubic metallic spinels $CuCr_2S_4$ and $CuCr_2Se_4$ [1] uncovered additional magnetic transitions significantly below their respective Curie temperatures of 376 K and 420 K [1–6]. These newly discovered transitions at 88 K for $CuCr_2S_4$ and 60 K for $CuCr_2Se_4$ indicate spin reorientations, i.e., the change of the spin structure from collinear to non-collinear at low temperatures. Preliminary neutron diffraction data on our $CuCr_2S_4$ samples indeed indicate that the ferromagnetic structure develops a small commensurate spin density wave or a conical distortion at low temperatures [7]. The observed temperature-dependent variations in electrical resistivity imply a transition from the mixed $Cr^{3+}/Cr^{4+}$ valency with collinear ferromagnetism at high temperatures to a charge-ordered state at low temperatures featuring a conical or more intricate structure, consistent with prior findings on $CuCr_2Se_4$ [8, 9]. Additional anomalies detected around 30 K and 15 K, respectively, are only weakly reflected in magnetization changes and may be associated with anisotropy effects.

The low-temperature magnetic transitions found in $CuCr_2S_4$ and $CuCr_2Se_4$ strongly resemble results on the isostructural $Fe_{1-x}Cu_xCr_2S_4$ system, in which Cu ions replace Fe in tetrahedral sites, and suggest a relation to Jahn-Teller instabilities found in the Cr system with higher Cu concentrations x [10, 11].

Previous studies on $CuCr_2S_4$ and $CuCr_2Se_4$ have demonstrated the significant impact of chemical substitutions on their magnetic and electronic properties by substituting the metal ions with diamagnetic ions. For instance, the magnetic phase diagram of $CuCr_{2-x}Ti_xS_4$ (Cu in tetrahedral sites and Cr/Ti in octahedral sites) deduced from magnetization measurement [12] suggests that chemical substitutions of chromium by titanium cause magnetic frustration, leading to a change in magnetic behaviour from ferromagnetic to spin-glassy. Further, the physical properties of $CuCr_2S_4$ have been interpreted with oxidation states $Cu^+Cr^{3+}Cr^{4+}S_4^{2-}$; thus, electrical properties and ferromagnetic behaviour were attributed to double exchange between $Cr^{3+}$ and $Cr^{4+}$ [13, 14]. Also, the magnetic properties of $CuCr_2Se_4$ have been shown to be sensitive to the substitution of Cr by Ti, revealing the coexistence of ferromagnetic and a spin-glassy regime [15]. $CuCr_{2-x}Sn_xSe_4$ showed evident ferromagnetic behaviour at x = 0.3, whereas at x = 0.7, antiferromagnetic (AF) behaviour was observed with a dominant ferromagnetic character in the exchange interactions [16]. Chemical substitutions of Cr, such as in $CuCr_{2-x}M_xX_4$ spinels (M=Zr, Sn, Ti, and X=S and Se), revealed that M is in the oxidation state 4+, whereas Cu exists as a diamagnetic $Cu^+$ cation [15–19].

To gain further information on the still insufficiently understood character of the recently [1] found low-temperature magnetic transitions in $CuCr_2S_4$ and $CuCr_2Se_4$, we investigated $CuCr_{2-x}Sn_xS_4$. With diamagnetic ions of the isotope $^{119}Sn$ substituting magnetic Cr ions in the octahedral sites of the compound, we could employ Mössbauer spectroscopy to study the temperature dependence of magnetic hyperfine fields at Sn in the ordered magnetic state. $^{119}Sn$ is known to reveal magnetic hyperfine fields in this compound due to supertransferred spin density by polarized covalent bonds from Cr to Sn via S [20–23]. For a coherent comparison with the undoped samples reported in Ref. [1], we have also performed magnetization and μSR on the Sn-doped samples. In contrast, the magnetic fields at the implanted muons in interstitial lattice sites are of magnetic dipolar origin and due to polarized conduction electrons. In this sense, the study of the magnetic hyperfine fields at $^{119}Sn$ is complementary to μSR since the mechanisms responsible for the measured fields are different. The aim was to see if the dramatic changes in local magnetic field and spin dynamics, as probed by the interstitially implanted μ$^+$ are also reflected at the substituted diamagnetic Sn.

To minimize magnetic dilution effects and keep the magnetic properties of the Sn doped samples as close as possible to the undoped one, the Sn concentrations x were chosen smaller than 0.1. The μSR measurements were conducted on $CuCr_{1.97}Sn_{0.03}S_4$. For Mössbauer spectroscopy a higher concentration of $^{119}Sn$ is required to obtain statistically significant spectra; hence, $CuCr_{1.92}Sn_{0.08}S_4$ was prepared with isotopically enriched $^{119}Sn$. Using a common sample for both techniques was not possible due to muon beam time schedule and synthesis limitations.

## 2. Experimental Details

Polycrystalline $CuCr_2S_4$, $CuCr_{1.97}Sn_{0.03}S_4$, and $CuCr_{1.92}Sn_{0.08}S_4$ were synthesized using solid-state reactions similar to the process described in Ref. [1]. High-purity elements of low oxygen Cu (99.9%, Alfa Aesar), Cr (99.999%, Alfa Aesar), Sn (99.99 Alfa Aesar, enriched to 90% with $^{119}Sn$ isotope), and S (99.999%, Puratronic), taken in stoichiometric ratio, were ground and mixed together in an argon box. The evacuated ampoules were heated at 600°C for 300 hours with two intermediary regrinding processes. As a final step, the polycrystalline product was annealed in a chalcogen-rich environment to minimize the possible anion defects.

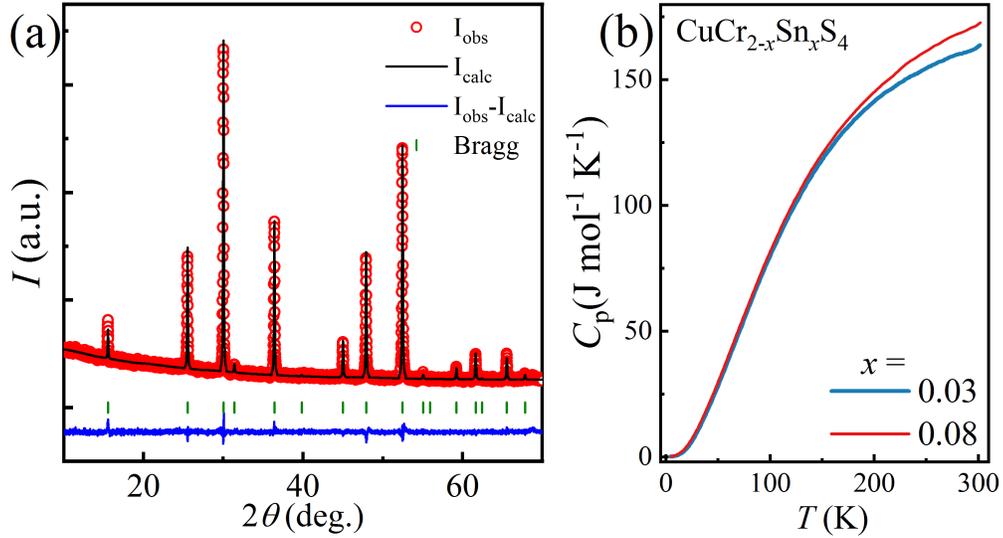

**Figure 1:** (a) X-ray powder diffraction pattern for polycrystalline $CuCr_{1.97}Sn_{0.03}S_4$. The measured intensities (red open circles) are compared with the calculated profile (black solid line). Bragg positions are indicated by vertical green bars. The difference between the observed and calculated intensities is shown by the blue line. (b) Temperature dependence of heat capacity $C_P$ measured for polycrystalline $CuCr_{1.97}Sn_{0.03}S_4$ and $CuCr_{1.92}Sn_{0.08}S_4$.

| Sample | $x_0$ (f.c.) | $a_0$ (Å) | $R_p$ | $R_{wp}$ | $R_{exp}$ | Chi$^2$ |
|---|---|---|---|---|---|---|
| $CuCr_2S_4$ | 0.2583(1) | 9.825(1) | 4.31 | 5.53 | 4.95 | 1.25 |
| $CuCr_{1.97}Sn_{0.03}S_4$ | 0.2583(1) | 9.841(1) | 3.65 | 4.70 | 4.15 | 1.28 |
| $CuCr_{1.92}Sn_{0.08}S_4$ | 0.2583(1) | 9.854(1) | 2.69 | 3.46 | 2.79 | 1.53 |

**Table 1:** Structural parameters of $CuCr_2S_4$, $CuCr_{1.97}Sn_{0.03}S_4$, and $CuCr_{1.92}Sn_{0.08}S_4$: lattice constants $a_0$, sulfur positional parameter $x_0$ (in fractional coordinates, f.c.), refinement parameters $R_P$, $R_{wp}$, $R_{exp}$ and goodness of fit Chi$^2$ obtained by Rietveld refinement of the x-ray patterns.

The phase purity was assessed through x-ray diffraction (XRD) using a STOE Stadi P diffractometer with CuKα radiation. Structural refinement was performed using the Rietveld method within the FullProfSuite software package [24].

Magnetic characterization was carried out using a commercial SQUID magnetometer (MPMS-3, Quantum Design) over a temperature range of 2 K and 400 K under magnetic fields up to 7 T. The heat capacity measurements were conducted on a thin rectangular-shaped bar using a Physical Properties Measurement System (PPMS, Quantum Design) within the 4 K to 300 K range. Muon spin rotation and relaxation (μSR) experiments were carried out on $CuCr_{1.97}Sn_{0.03}S_4$ at the Swiss Muon Source, Paul Scherrer Institute, Switzerland. A 100% spin-polarized positive muon (μ$^+$) beam was employed using the πM3 beamline of the General Purpose Surface-Muon instrument (GPS). Zero field (ZF) measurements on the polycrystalline sample over a temperature range of 5 K to 220 K (below $T_C$). The sample was enclosed in aluminized Mylar foil and mounted on an ultrapure copper fork within a helium flow cryostat to ensure optimal thermal stability.

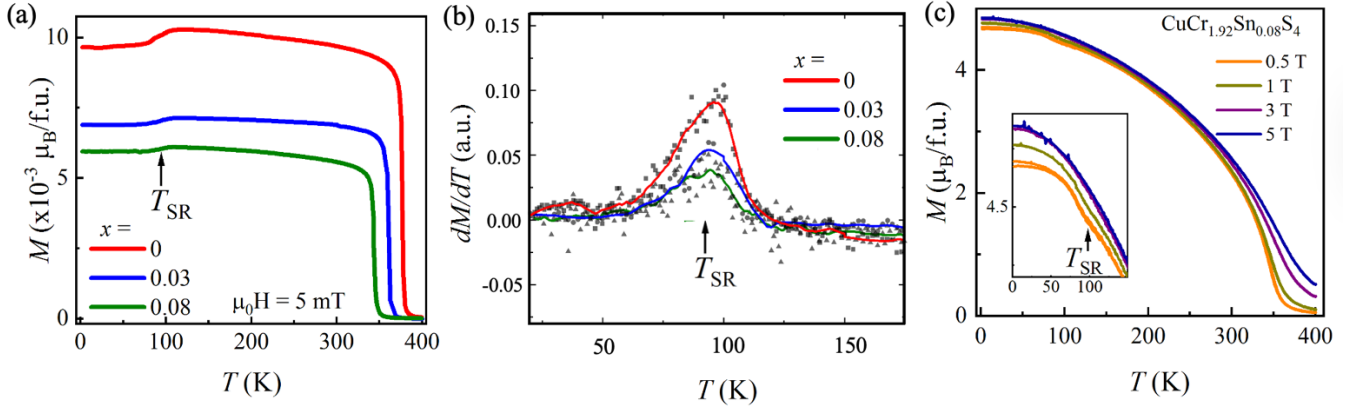

**Figure 2:** (a) Temperature dependences of field-cooled (FC) magnetization of pure $CuCr_2S_4$ and Sn-doped $CuCr_{1.97}Sn_{0.03}S_4$ and $CuCr_{1.92}Sn_{0.08}S_4$ measured in a magnetic field of 5 mT. (b) Temperature dependences of the first derivative $dM/dT$ for different concentrations of $x = 0$, 0.03, and 0.08. (c) Temperature dependences of magnetization of $CuCr_{1.92}Sn_{0.08}S_4$ measured in magnetic fields 0.5-5 T. Inset in (c) shows a zoom-in of the magnetization at the spin-reorientation $T_{SR}$. The vertical arrows mark the spin-reorientation temperature.

Its placement was carefully adjusted within the muon beam path, and the efficient thermal contact was maintained via helium exchange gas.
The 'VETO' mode was activated to suppress the background signal from muons stopping outside the sample. The analysis of spectra was performed using the musrFit software package [25]. Details on the experimental method and theoretical principles of μSR can be found in the literature (see e.g., [26–28]).

Details on $^{119}$Sn Mössbauer spectroscopy may be found, e.g., in [29]. The spectra of $CuCr_{1.92}Sn_{0.08}S_4$ were taken at Brasileiro de Pesquisas Físicas (CBPF), Brazil, using a Montana variable temperature closed-cycle cryostat in the temperature range from 3 to 300 K. As radioactive source (5 mCi) served $^{119m}$Sn:CaSnO$_3$ kept at room temperature, moved in sinusoidal mode outside of the cryostat. Annotations to the calibration of the Doppler velocity are given in the Supplement. The center shifts of spectra (isomer shifts) refer to the source material $^{119m}$Sn:CaSnO$_3$. A Kr proportional counter was used as a detector for the 23.8 keV gamma ray resonance transition. Data analysis was performed using the Mosswinn 4.0i software [30].

## 3. Results
### A. Sample characterization

To characterize the phase purity and phase transitions of synthesized $CuCr_{1.97}Sn_{0.03}S_4$ and $CuCr_{1.92}Sn_{0.08}S_4$, we investigated the structure, specific heat $C_P$, magnetic susceptibility ($\chi$), and magnetization ($M$) and compared them with the pure $CuCr_2S_4$. The x-ray powder diffraction pattern of Sn-doped $CuCr_2S_4$ corresponds to a single phase $Fd\text{-}3m$ symmetry of spinel structure, Figure 1(a). The calculated structural parameters, such as lattice parameters, fractional coordinates of sulfur, and refinement parameters that show the goodness of fit, are summarized in Table 1. Rietveld refinement of $CuCr_{1.97}Sn_{0.03}S_4$ and $CuCr_{1.92}Sn_{0.08}S_4$ reproduced the experimental intensities very well, and the results show a clear increase of lattice parameters, $a_0$, compared to the pure $CuCr_2S_4$.

The heat capacity of $CuCr_{1.97}Sn_{0.03}S_4$ and $CuCr_{1.92}Sn_{0.08}S_4$ exhibits a typical increase with rising temperature and shows no indication of phase transitions within the measured temperature range of 4 K to 300 K (see Figure 1(b)). Figures 2(a) show the temperature dependence of magnetization measured for pure $CuCr_2S_4$ and compared with $CuCr_{1.97}Sn_{0.03}S_4$ and $CuCr_{1.92}Sn_{0.08}S_4$.

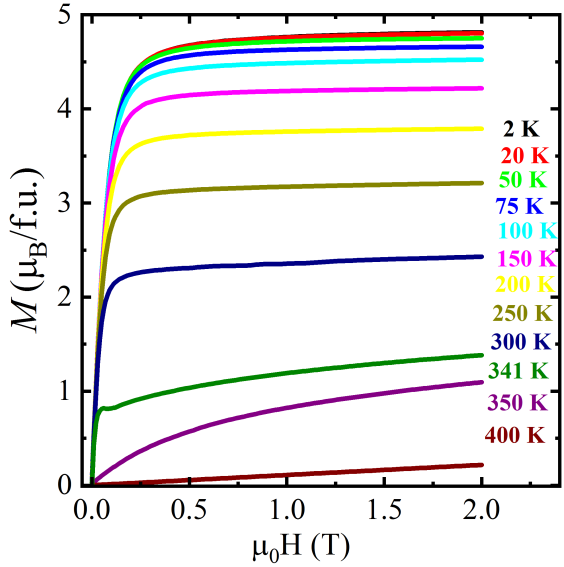

**Figure 3:** Field-dependent magnetization curves measured at different temperatures for polycrystalline $CuCr_{1.92}Sn_{0.08}S_4$.

The low-field (5 mT) magnetization data show that the substitution of Cr by Sn at the octahedral sites reduces the ferromagnetic ordering temperature from $T_C$ = 377 K to 360 K in the case of $CuCr_{1.97}Sn_{0.03}S_4$ and 343 K for $CuCr_{1.92}Sn_{0.08}S_4$. The critical temperature $T_C$ was calculated from the minimum of the first derivative $dM/dT$.

Upon cooling, the magnetization curves show a downturn at approximately 100 K, which is similar to the value observed for pure $CuCr_2S_4$ [1] (there labelled as $T_2$). The first derivative $dM/dT$ exhibits a maximum at approximately 88 K (see Figure 2(b)), which is associated with the spin reorientation transition $T_{SR}$ [1]. The value of $T_{SR}$ remains nearly unchanged upon increasing the Sn concentration up to x=0.08. At first glance, the anomaly in $dM/dT$ appears to be suppressed with increasing Sn content. For higher magnetic fields, the magnetization curves of $CuCr_{1.97}Sn_{0.03}S_4$ and $CuCr_{1.92}Sn_{0.08}S_4$ are shifted to higher temperatures. Interestingly, the anomaly observed at $T_{SR}$ in a field of 5 mT also remains detectable at a field of 1 T but completely vanishes in fields higher than 3 T (see Figure 2(c)).

The field-dependent magnetization of Sn-substituted $CuCr_2S_4$ shows a typical ferromagnetic behaviour (Figure 3). At 2 K, the calculated magnetic moment in 5 T is about 4.6 $\mu_B$/f.u., which is slightly lower than 4.7 $\mu_B$/f.u. for pure $CuCr_2S_4$.

## B. Muon spin rotation and relaxation

μSR allows investigation of internal magnetic fields at the atomic scale. In this technique, spin-polarized muons are implanted into a sample, where they are stopped at interstitial lattice sites. The interaction with local magnetic fields $B_\mu$ will lead to the precession of muon spins with frequency $\nu_\mu = \gamma_\mu/(2\pi) B_\mu$ ( $\gamma_\mu$=2π×135.5 μs$^{-1}$T$^{-1}$ is the muon gyromagnetic ratio). In the present case, the actual stopping sites are not known. An assignment of the possible muon stopping sites via ab initio calculations is not yet available for this system. Based on electrostatic and crystallographic arguments and in analogy to related spinel systems [31], muons are likely to occupy interstitial positions near sulphur anions or between sulphur-sulphur pairs, where the local electrostatic potential minima occur. The local fields acting on the muon spin are due to various contributions from dipolar fields from nearby Cr magnetic moments, the Lorentz field from domain magnetization and contact hyperfine fields caused by polarized conduction electrons [32].

In the experiment, the time evolution of the asymmetry function A(t) of the decay positron count rates was analyzed to extract the time-dependent polarization of the muons following implantation at t=0: A(t)/A(t=0)=$G_z$(t).

The zero field μSR (ZF-μSR) spectra of $CuCr_{1.97}Sn_{0.03}S_4$ at several temperatures are shown in Figures 4 and 5 for both short- and long-time windows. The spectra are close in appearance to those reported for the undoped compound with one oscillatory component above 80 K and at least two below [1]. The analysis was conducted using a three-component model (oscillatory (OSC), Lorentzian Kubo-Toyabe (LKT), and background (BG) terms), following the same fit strategy as used and described in [1]:

$$A_0 G_{LRO}^{ZF}(t) = A G_{osc}^{ZF}(t) + A_{BG} \quad (1)$$

for temperatures between $T_{SR}$ and $T_C$, and

$$A_0 G_{LRO}^{ZF}(t) = A G_{osc}^{ZF}(t) + A_{LKT} G_{LKT}^{ZF}(t) + A_{BG} \quad (2)$$

for temperatures below $T_{SR}$. Here, the oscillatory component is expressed as:

$$G_{osc}^{ZF}(t) = \sum_{i=1}^{2} A_i \left[ \frac{2}{3} e^{-\lambda_{T,i} t} \cos(2\pi \nu_{\mu,i} t) + \frac{1}{3} e^{-\lambda_{L,i} t} \right] \quad (3)$$

and the Lorentzian Kubo-Toyabe function is given by:

$$G_{LKT}^{ZF}(t) = \frac{2}{3} e^{-a_{LKT} t}(1 - a_{LKT} t) + \frac{1}{3}. \quad (4)$$

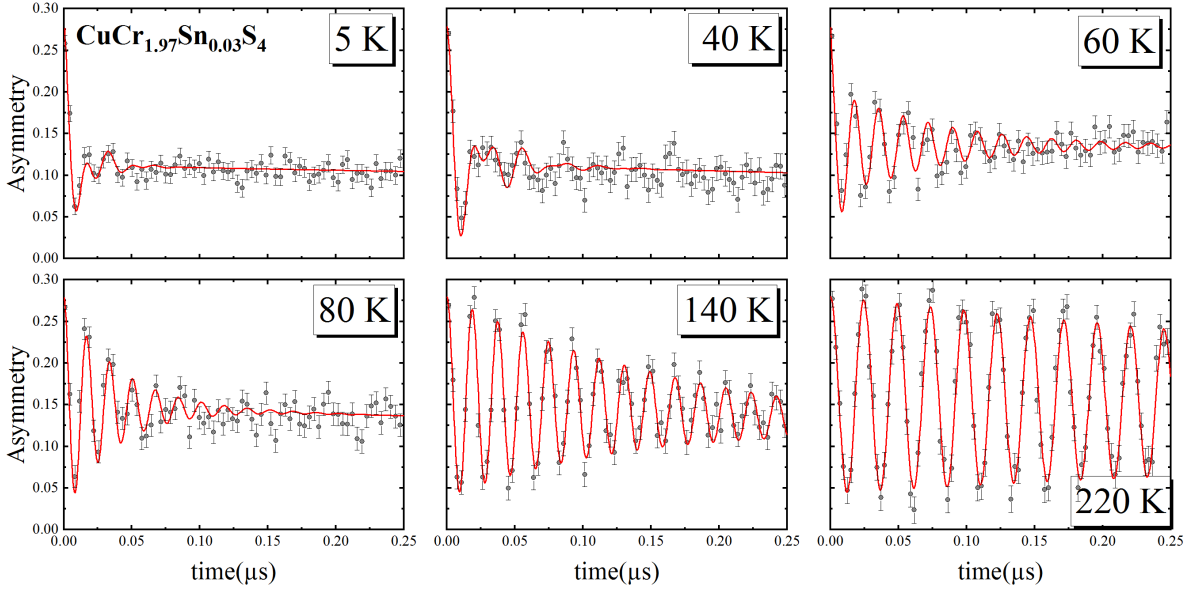

**Figure 4:** ZF-μSR spectra of CuCr$_{1.97}$Sn$_{0.03}$S$_4$ recorded at selected temperatures (5 K - 220 K) within a 0 - 0.25 μs time window, illustrating spontaneous muon spin rotation in the magnetically ordered state. The solid lines represent fits to the data based on equations 1 and 2.

The oscillatory term $G_{osc}^{ZF}(t)$ represents the sum of signals associated with the muon spin precession in local fields generated by ordered electronic magnetic moments. The muon spin precesses with rotation frequencies $\nu_{\mu,i}$ as seen in the first term in the square bracket of eq. 3. The damping parameters $\lambda_{T,i}$ describe the transverse relaxation, which arises primarily from the field inhomogeneities. The second term in the oscillatory expression, known as "1/3 tail", corresponds to the field dynamic fluctuations along with the initial muon polarization direction, characterized by the longitudinal damping rate.

At temperatures of 80 K and below, an additional time-dependent contribution with reduced asymmetry was necessary to obtain an acceptable fit. This was accomplished by incorporating a so-called Lorentzian Kubo-Toyabe function $G_{LKT}^{ZF}(t)$, which describes the spectral response of randomly distributed local fields at the muon sites with an average value of zero. While a Gaussian field distribution is typically expected in concentrated systems with randomly oriented static dipoles, the present data rather suggests a distribution of

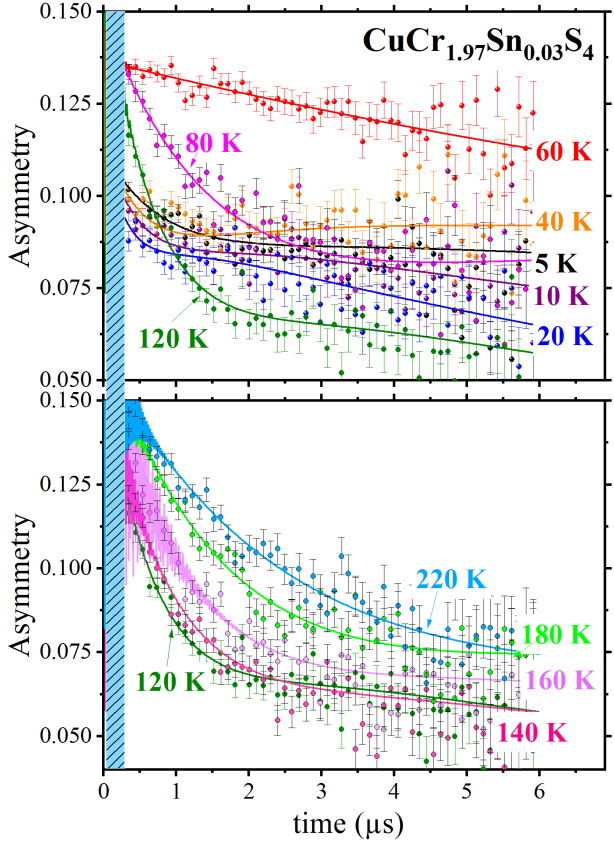

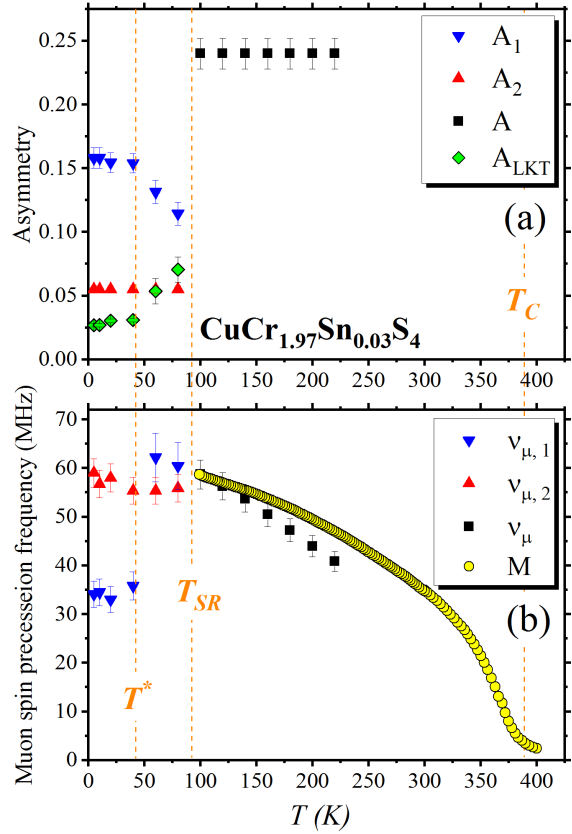

**Figure 5:** ZF-μSR spectra of CuCr$_{1.97}$Sn$_{0.03}$S$_4$ displayed with lower time resolution, extending up to 6 μs, illustrating the dynamic relaxation behavior at selected temperatures. Panel (a) corresponds to T < 120 K, while panel (b) shows the data for T > 120 K in (b). The highlighted blue rectangle in Figure 4 represents a zoomed-in section of the shorter time within the magnetically ordered state. Solid curves correspond to the fitted data obtained using equations 1 and 2.

**Figure 6:** Temperature dependence of ZF μSR parameters in CuCr$_{1.97}$Sn$_{0.03}$S$_4$: (a) Asymmetries; A, A$_i$, A$_{LKT}$, (b) spontaneous μ$^+$ spin precession frequencies; ν and ν$_{μ,i}$. Open symbols for M in (b) are the magnetization data measured in an external field of 1 T normalized to the value of μSR rotation frequencies at 100 K. Dashed lines indicate the magnetic phase transitions identified in magnetization measurements.

Lorentzian shape similar to that found in systems with dilute dipole moments.

The decay rate $a_{LKT}$ of polarization is proportional to the width of the field distribution [27, 33], $a_{LKT}/\gamma_\mu$ defining the half-width at half maximum.

Initially, the muon spin polarization undergoes an exponential decay, followed by a minimum, and subsequently stabilizes at 1/3 and stays time-independent.

The background asymmetry A$_{BG}$ originates from a non-magnetic signal, likely from the sample holder. A$_{BG}$ was determined from a measurement in a weak transverse field of 5 mT and was kept fixed at 0.025 for the ZF-μSR data fit.

The ZF-μSR spectra between 5 K and $T_C$ exhibit distinct oscillations at the early times (seen at high resolution), indicative of a magnetically ordered state (see Figure 4). Three distinct temperature regimes can be identified. The spectra in the temperature range between $T_{SR}$ and $T_C$ can be well described using equation 1, with a single oscillation frequency.

In the intermediate temperature range between $T^*$ < T < $T_{SR}$, the oscillation frequency exhibits noticeable variation, along with corresponding changes in the relaxation parameters. Below $T^* \sim$ 40 K (referred to $T_1$ in [1]), the ZF-μSR spectra become more complex, showing strongly damped oscillatory signals at multiple frequencies, which are well reproduced by eq. 2.

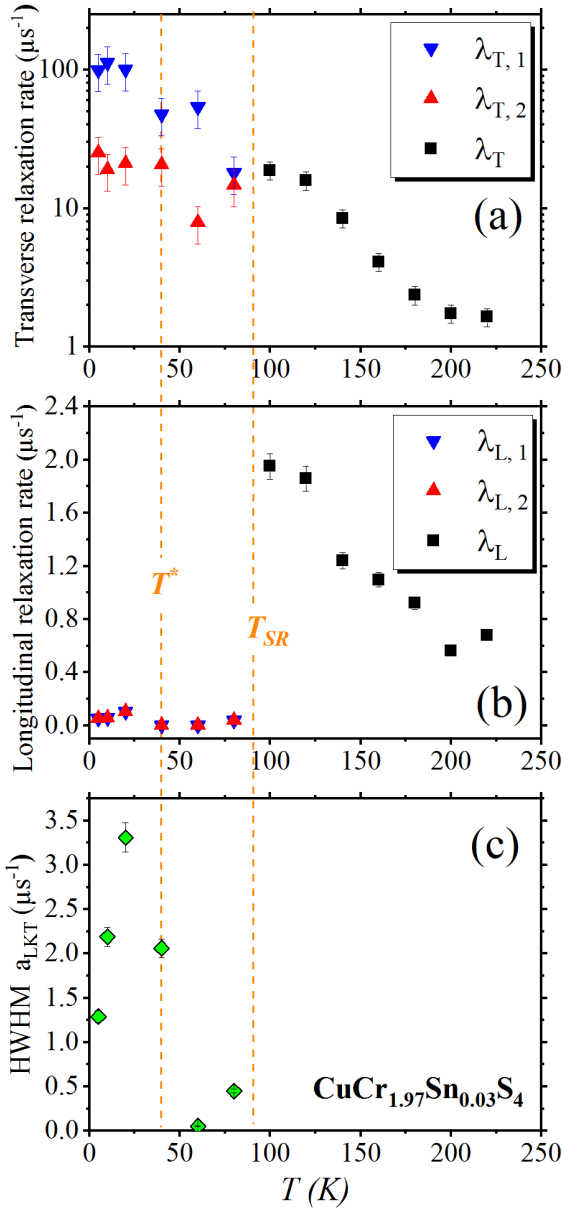

**Figure 7:** Temperature dependence of ZF-μSR parameters in CuCr$_{1.97}$Sn$_{0.03}$S$_4$: (a) transverse relaxation rates $\lambda_T$ and $\lambda_{T,i}$, (b) longitudinal relaxation rate $\lambda_L$ and $\lambda_{L,i}$, (c) half-width at half maximum of field distribution of Kubo-Toyabe component. The dashed lines indicate the magnetic phase transition identified in magnetization measurements.

Figure 6(a) illustrates the evolution of the asymmetry parameters for the oscillatory and relaxation components. Above $T_{SR}$, the asymmetry is mainly attributed to a single oscillatory component (A) and a background signal (A$_{BG}$). Below $T_{SR}$, the asymmetry amplitude A splits into two oscillatory signals (with amplitudes A$_1$ and A$_2$), along with an additional Lorentzian Kubo-Toyabe component (A$_{LKT}$), which means that the muons are stopping in different magnetic environments. The temperature-dependent variation of the frequency $\nu_\mu$ (Figure 6(b)) follows a magnetization-like trend down to $T_{SR}$. Below $T_{SR}$, the oscillation frequencies $\nu_{\mu,1}$ increases up to 65 MHz between $T_{SR}$ and $T^*$ before dropping to approximately 35 MHz below $T^*$, while $\nu_{\mu,2}$ remains nearly constant at 57 MHz down to 5 K. The transition is accompanied by substantial changes in both transverse and longitudinal relaxation rates, as illustrated in Figures 7(a) and 7(b).

The temperature-dependent evolution of the longitudinal relaxation rates ($\lambda_L$ and $\lambda_{L,i}$) and the half width at half maximum (HWHM a$_{LKT}$) of the field distribution of the Kubo-Toyabe component are presented in Figures 7(b) and 7(c). The values of $\lambda_{L,1}$ and $\lambda_{L,2}$ exhibit an increase near $T_{SR}$, coinciding with the rise in transverse damping $\lambda_T$. Additionally, the HWHM a$_{LKT}$ parameter (Figure 7(c)) displays a pronounced increase below $T^*$.

### C. Mössbauer spectroscopy

$^{119}$Sn Mössbauer spectra on CuCr$_{2-x}$Sn$_x$S$_4$ have been reported several times [20–22]. Though being itself diamagnetic, Sn can serve as a sensitive probe for magnetism via directly or indirectly transferred magnetic hyperfine fields. As shown in [21, 23], it is dominantly supertransferred magnetic hyperfine fields (B$_{sthf}$)

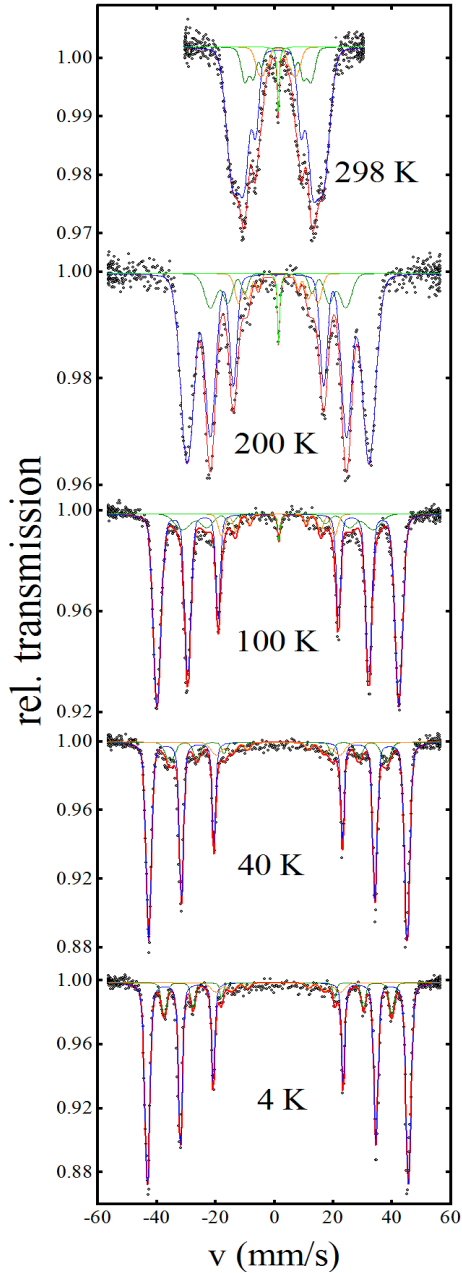

**Figure 8:** $^{119}$Sn Mössbauer spectra of CuCr$_{1.92}$Sn$_{0.08}$S$_4$ at various temperatures.

acting on Sn in CuCr$_{2-x}$Sn$_x$S$_4$. Sn is known to be substituted for Cr in the octahedral sites of the spinel lattice.

In the ferromagnetic state, the t$_{2g}$ orbitals of Cr$^{3+}$ become spin-polarized. Covalent mixing leads to spin polarization of p orbitals of the neighbouring S$^{2-}$, which in turn transfers spin density into Sn$^{4+}$

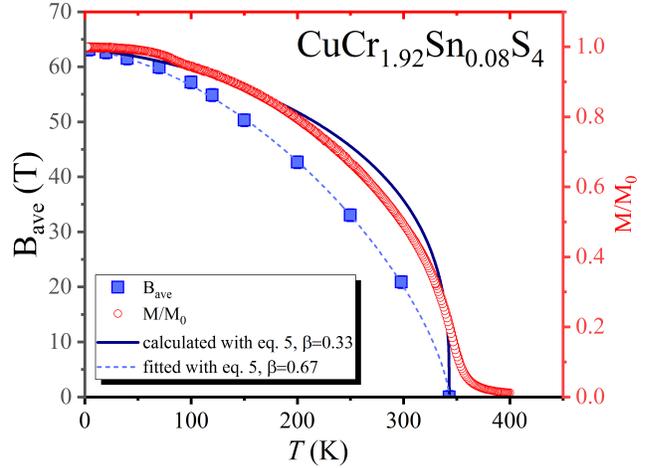

**Figure 9:** Temperature dependence of averaged magnetic hyperfine fields at $^{119}$Sn in CuCr$_{1.92}$Sn$_{0.08}$S$_4$ (dashed line is the fit to equation 5) and normalized magnetization ($\mu_0 H$ = 0.5 T) compared with a calculated dependence, using equation 5.

5s orbitals that create a magnetic field at the Sn nuclei via Fermi contact interaction.

The reported value of B$_{sthf}$ found for CuCr$_{1.95}$Sn$_{0.05}$S$_4$ at 4 K is very large, with about 58 T, and opens the possibility of following its temperature dependence over a wide range, as demonstrated by [23].

These earlier data were collected with a relatively low degree of statistics and revealed strongly broadened resonance line widths even at low temperatures [20, 22, 23].

As seen from the spectra shown in Figure 8, obtained from the present preparation for CuCr$_{1.92}$Sn$_{0.08}$S$_4$, this problem could be overcome with our samples. The line widths (FWHM) at 4 K are narrow, with about 1 mm/s. The magnetic hyperfine splitting corresponds to B$_1$= 66(2) T, which is even larger than the earlier reported values.

The other hyperfine parameters, nearly vanishing nuclear electric quadrupole interaction as expected for highly symmetric octahedral sulfur coordination and an isomer shift of 1.4 mm/s with respect to the $^{119m}$Sn:CaSnO$_3$ source at room temperature, indicating a rather covalent bonding

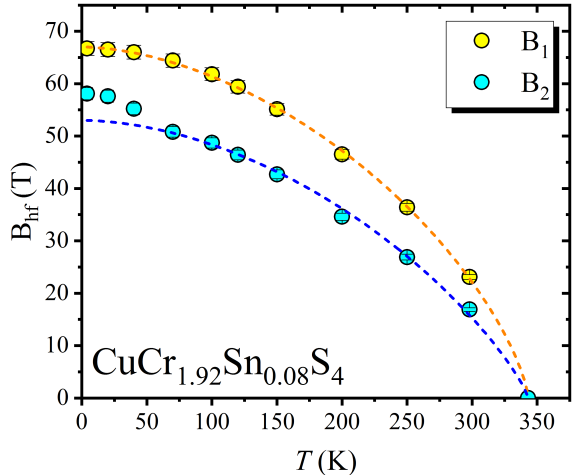

**Figure 10:** Temperature dependence of $^{119}$Sn magnetic hyperfine fields $B_1$ and $B_2$. The dashed lines are fits to equation 5. For details see text.

of formally tetravalent tin, are in agreement with the literature [23].

In the spectra taken at 100 K and above, a single resonance line becomes visible in the center of the spectra with an isomer shift close to that of the main compound. It is indicative of a paramagnetic impurity that becomes magnetically ordered at lower temperatures. Its relative contribution is around 1 - 2 % of the total spectral area and cannot be resolved when magnetically split at lower temperatures. A possible compound may be $Cr_2S_3$.

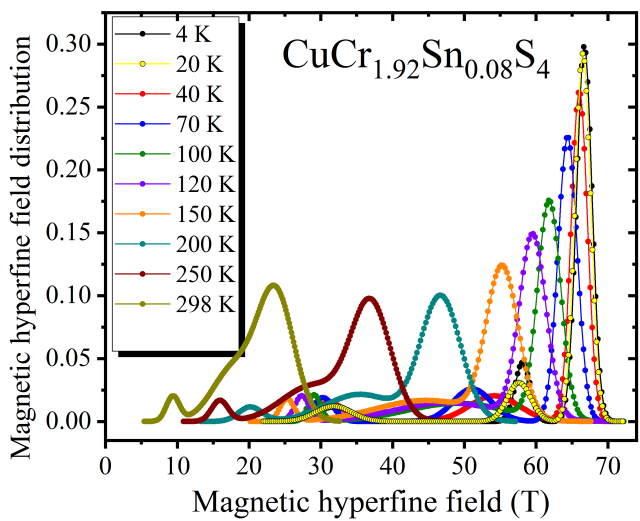

**Figure 11:** Distribution of magnetic hyperfine fields of $^{119}$Sn in $CuCr_{1.92}Sn_{0.08}S_4$ at various temperatures.

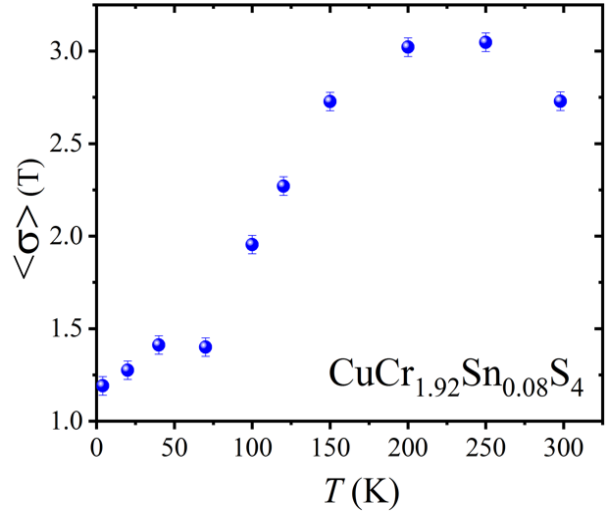

**Figure 12:** Temperature dependence of the weighted average of the Gaussian width of $^{119}$Sn magnetic hyperfine field distribution.

For the fits of spectra below 100 K, this contribution will be neglected.

In addition to the major intensity line sextet, the improved spectral resolution allows to reveal the presence of a second sextet with smaller magnetic splitting corresponding to $B_2$ and an area of about 1/5 of that of the major sextet that could not be resolved in earlier experiments. Within fit uncertainty, the isomer shift is the same as for the main sextet.

An analysis allowing for several magnetic sextet spectra with differing mean hyperfine fields and an assumed Gaussian-shaped distribution profile suggests even the presence of a third component with a yet still lower spectral contribution. The relative spectral areas of these subspectra were kept fixed for the various temperatures to the values derived at 4 K. Fitted selected spectra are shown in Figure 8. They reveal a decreasing magnetic splitting upon rising temperature. Figure 9 shows the temperature dependence of the averaged magnetic hyperfine fields $B_{ave}$. The average was done by weighting the field values of fitted subspectra by the relative areas of the corresponding subspectra. We have chosen this plot for a better comparison with earlier data [23], where the additional subspectra could not be resolved.

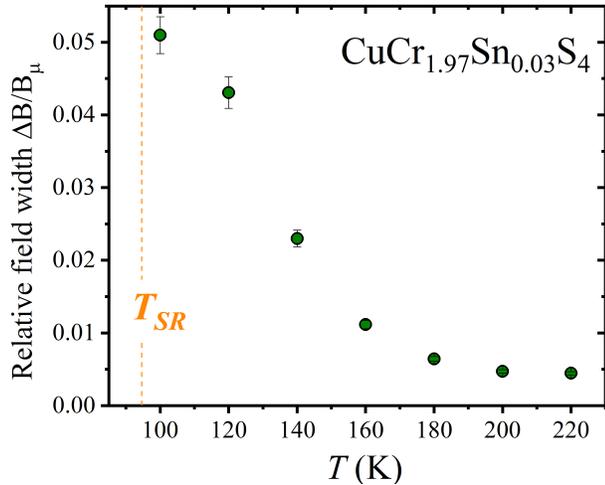

**Figure 13:** Dependence of the temperature of the relative field distribution for CuCr$_{1.97}$Sn$_{0.03}$S$_4$ from µSR rotation signals.

A separate plot of $B_1(T)$ and $B_2(T)$ is shown in Figure 10. Figure 11 presents the corresponding development of the distribution of magnetic hyperfine fields. As can also be observed directly from the change of spectral shapes, the distributions reveal an onset of strong broadenings between 70 and 100 K, i.e., close to the transition traced from µSR (see Figure 12).

### 4. Discussion

Our magnetization studies indicate that Sn doping leads to a decrease in Curie temperature and a reduction of magnetic moment per formula unit, as expected from the dilution in the Cr sublattice by diamagnetic Sn. Again, we can trace transitions at low temperatures. The spin reorientation observed around 88 K in the undoped material [1] is shifted to $T_{SR}$ = 80 K. The transition associated in Ref. [1] (there labelled as $T_1$) with a change in anisotropy at $T^*$ is less clearly expressed in the Sn-doped samples. This demonstrates that Sn doping at the percent level only leads to magnetic dilution effects. We thus can expect that the µSR and Mössbauer data may serve for proper comparison. The µSR spectra are indeed comparable to those reported for the undoped material [1]. There are some differences in the relative asymmetry contributions of sub-patterns and their temperature dependence in the low-temperature phase, representing magnetically different neighbourhoods to the stopping sites. This may be caused by differences in dipolar fields due to nearby diamagnetic Sn and/or the presence of the magnetic impurity. Yet the major qualitative changes in a local field and damping related to the magnetic transitions are the same. A still ununderstood feature is the origin of the Kubo-Toyabe contribution found at low temperatures. It has only low spectral weight and has been introduced purely phenomenologically to improve fit quality without affecting the major conclusions related to magnetic transitions. For the moment, we can only speculate about a possible spin-glassy contribution within the ferromagnetic phase due to local frustration within the Cr pyrochlore-type sublattice with next-nearest antiferromagnetic exchange interactions.

In analogy to the data on the undoped material [1], the spontaneous rotation $\nu_\mu$ shows a steeper temperature dependence above 100 K when compared to magnetization. The local field $B_\mu$ is a complex superposition of several, in part counteracting, contributions. While the Lorentz field should vary proportional to magnetization, the superposition of dipole fields from neighboring moments and the contact fields by polarized conduction electrons can lead to a temperature dependence deviating from that of magnetization. The appearance of two distinct frequencies at lower temperatures should be related to changes in local dipolar contributions at different stopping sites due to spin reorientation, while contact contributions are, in first approximation, isotropic. The Lorentz field (corresponding to about 20 MHz) should be the same for both sites, and according to the relatively small changes of magnetization around $T_{SR}$, it is supposed to undergo only a corresponding minor variation upon the reorientation transition.

The present Mössbauer spectra of $CuCr_{1.92}Sn_{0.08}S_4$ are compatible with those from earlier studies [23] yet were taken with improved statistics. The essential additional information from these new data is the resolution of at least one further sextet (sub)spectrum with a lower spectral area and a smaller magnetic hyperfine field $B_2$. The narrow linewidths of the spectra at low temperature turn broadened above about 80 K. One possible scenario could be that the less weighted spectrum is related to Sn having another Sn substituting a Cr within its six cation octahedral site neighbours (in analogy, this would hold for the eventual only poorly resolved third sextet pattern). From estimated ratios of spectral areas using simple binomial occupation based on the nominal Sn doping concentration among nearest neighbour octahedral sites, this assumption is compatible with the fits from spectra and, thus, is a possible explanation. The reason for the smaller magnetic field values $B_2$ for the lower weight sextet can then be related to the reduced number of nearest neighbour Cr ions. In the fits, the relative spectral contributions from the sextet spectra could be kept unchanged for varying temperatures, indicating that the spectra may indeed be related to Sn sites with different numbers of magnetic neighbours on octahedral sites.

As shown in Figure 9, $B_{ave}$ (T) reveals a steeper behaviour above 100 K when compared with a calculated temperature dependence using the generic formula

$$\frac{B(T)}{B(T=0)} = \left[1 - \left(\frac{T}{T_C}\right)^\alpha\right]^\beta \qquad (5)$$

with α = 3/2, β = 1/3. These values are those expected for low-temperature spin wave excitations and for the approach to the Curie temperature (here $T_C$ = 343 K) of a 3D ferromagnet. A fit of $B_{ave}$ using equation 5 gives α=1.5, β ≈ 0.67, and B(T=0) = 63 T. The more detailed plot of Figure 10 shows that $B_2$(T) tends to saturate below 100 K with an indication of a weak additional rise below 70 K. For a better visualization, we have included a fit (equation 5) to $B_2$(T) between high temperatures and 100 K, extrapolating the fit to T=0 (α=1.8, β ≈ 0.81, and an extrapolated B(T=0) = 53 T). This additional increase resembles the variation of magnetization near $T_{SR}$, yet there is no visible change in $B_1$(T). For $B_1$(T) we get α=1.7, β=0.69 and B(T=0) = 67 T.

Also, the earlier, less resolved spectra of [20, 21, 23] showed similar variations of the magnetic hyperfine field with temperature, with a clearly steeper dependence than that expected for a calculated one using a Brillouin law for S=3/2. For its interpretation, an adopted model by Coey and Sawatzky [34] related to disorder in local exchange interactions was used, which used a local molecular field theory to study the changes in magnetic hyperfine spectra of magnetic iron compounds upon substituting non-magnetic ions. They had assumed a dominant superexchange mechanism between iron and its nearest neighbour cations. The non-magnetic substitution causes a local distribution of the molecular field, leading to a steeper temperature dependence hyperfine field depending on the number of non-magnetic neighbours of the Mössbauer atom. In addition, distributions of hyperfine fields are predicted.

We can apply the above-sketched arguments also in the present case with the now at lowest temperatures resolved magnetic spectra due to hyperfine fields transferred to non-magnetic Sn in magnetic surrounding with varying numbers of magnetic and non-magnetic neighbour cations.

The magnetic hyperfine fields at the substituted Sn at Cr sites are assumed to be dominantly of a supertransferred nature due to Cr via polarized S ions. The difference in the values of Sn hyperfine fields $B_1 = 67$ T and $B_2 = 58$ T associated with Sn having 6 and 5 nearest neighbouring Cr (+ 1 Sn) measured at 4 K suggest that the transferred field per Cr neighbour amounts to about 9 T being supplemented by a common additional field of about 13 T due to other mechanisms like conduction electron polarization.

The slightly steeper temperature dependence of $B_2(T)$ compared with $B_1(T)$, both being steeper than the calculated dependence for a 3D magnet, is in good qualitative agreement with the model of Coey and Sawatzky predicting for a larger number of non-magnetic (in the present case Sn) neighbors a steeper temperature dependence. Also, the observed hyperfine field distribution is in agreement with this model.

There remains, however, a major question: How compatible are the results drawn from the Mössbauer data with those from µSR? The reason why the magnetic transition around 80 K is clearly marked by the µSR data but (apart of the small anomaly in the temperature dependence of $B_2(T)$) are not visible in the magnetic hyperfine field at Sn substituted in the site of magnetic Cr must be sought in the different mechanisms how these fields are created. As shown in [1], the major changes in the fields sensed by interstitial $\mu^+$ arise from changes in dipolar fields from neighboring Cr moments. The corresponding dipolar fields at Sn can be estimated to be only on the order of some tenths of Tesla, i.e., are much smaller than the large supertransferred fields, and thus changes in these dipolar contributions cannot be traced. Supertransferred fields by themselves may, however, reflect spin canting as suspected in the magnetic structure below 70 K. This has, e.g., been demonstrated for a canted spin structure in Sn-doped iron garnets [35]. In our case, we may propose a highly symmetric nearest neighbour spiral configuration, with field components perpendicular to the major ferromagnetic component that may mutually cancel at the site of Sn (a possible case for $B_1$). If one of the octahedral neighbors is missing (case for $B_2$), there will, however, stay a residual perpendicular contribution, adding to the reduced ferromagnetic component, leading to the increase of $B_2$ below 70 K. Thus, the compatibility of µSR and Mössbauer results appears to be given.

We should, however, briefly consider another possible scenario in which the Sn subspectra with different magnetic hyperfine fields are related to different numbers of $Cr^{3+}$ (spin S=3/2) and $Cr^{4+}$ (S=1) as nearest cation neighbors of Sn. This would be expected for a static mixed-valent state at low temperatures with a relatively low contribution from $Cr^{4+}$. In this picture, the observed broadening of the Mössbauer spectra and the steeper behavior of B(T) above 70 K may be due to increased charge and related spin dynamics. Also, from µSR, an unusual variation of both transverse and longitudinal damping data (Figures 7 (a) and (b)) is observed. In Figure 13, we show the development of relative field width $\Delta B/B_\mu$ (i.e., the ratio between transverse damping rate $\lambda_T$ and spontaneous rotation frequency $\nu_\mu$). Its decrease above 80 K clearly indicates a variation in inhomogeneous broadening $\Delta B$, but in parallel, there is also a reduction of longitudinal damping rates $\lambda_L$ (see Figure 7(b)). Both µSR and Mössbauer data are thus indicative of a dynamic onset of a precursor to the spin-reoriented phase found below 70 K. A possible model could be the gradual onset of fluctuating moment components of Cr perpendicular to the axial ferromagnetic that become static below 70 K forming a conical spin structure for the Cr sublattice as supported by the preliminary neutron scattering results [7]. Notably, longitudinal damping $\lambda_L$ is nearly vanishing in the low-temperature phase (see Figure 7 (b)), i.e., spin fluctuations drop to the static limit. Taking into account the µSR data, the observed broadening of linewidths in the Mössbauer spectra above 70 K may thus at least partially be caused by spin dynamics instead or in addition to the static inhomogeneous broadening considered in the above-described first scenario.

## 5. Conclusions

Magnetization, µSR, and $^{119}$Sn Mössbauer spectroscopy performed on $CuCr_{2-x}Sn_xS_4$ have shown that the recently reported spin re-orientation transitions for $CuCr_2S_4$ and $CuCr_2Se_4$ are also found in Sn-doped samples.

While µSR reveals distinct changes in muon spin rotation frequencies and their damping around $T_{SR}$ = 80 K and $T^*$= 40 K, the changes observed in the Mössbauer hyperfine parameters are less pronounced. We could show that this different response can be understood from the different sites of the local probes; positive muons are stopped after implantation on interstitial sites, while Sn is substituted for Cr on octahedral sites of the spinel lattice. In addition, the magnetic fields sensed by the muons and the Sn Mössbauer atoms are related to different magnetic coupling mechanisms: Muon spins are dominantly subject to dipolar fields and their changes upon spin re-orientation from nearest neighbour magnetic Cr ions, while the fields at $^{119}$Sn are largely of supertransferred nature via spin-polarized p shells of neighbouring Sulphur cations.

Below $T_{SR,}$ the Mössbauer spectra reveal a resolved superposition of magnetic hyperfine patterns with different resonant absorption areas, indicating Sn sites typical for distinct magnetic surroundings that can be interpreted either with finite probabilities of substitutions of non-magnetic Sn on neighbouring Cr sites or alternatively with the partial presence of $Cr^{4+}$ within a dominant $Cr^{3+}$ neighbourhood. Upon rising temperatures above 70 K, the magnetic hyperfine splitting shows broadenings that can be explained by local magnetic static disorder introduced by the doping of diamagnetic Sn into the magnetic Cr lattice and/or by relaxational broadening due to increasing charge and spin fluctuations.

Similarly, the changes in transverse and longitudinal relaxation found by µSR above 80 K support the appearance of a precursor phase leading to the proposed low-temperature conical or spin-density wave phase. The detailed nature of the magnetic ground state is subject to ongoing neutron scattering and x-ray diffraction projects.

## Acknowledgements

E.S. expresses gratitude for the support received through the MARIA REICHE POSTDOCTORAL FELLOWSHIPS at TU Dresden, funded by the Free-state of Saxony and the Federal and State Program for Women Professors (Professorinnenprogramm). Additional funding was provided by the Deutsche Forschungsgemeinschaft via SFB 1143 (project-id 247310070), SFB 1415 (project-id 417590517), and the Würzburg-Dresden Cluster of Excellence on Complexity and Topology in Quantum Matter—ct.qmat (EXC 2147, project-id 390858490).

L.P. and V.T. acknowledge funding from the ANCD project (code 011201) in Moldova. E.M.B.S. appreciates the support from CNPq (grant 308756/2021-5) and multiple grants from FAPERJ, including a Professor Emeritus Fellowship (E-26/201.622/2021). C.P.C.M. and M.A.V.H. acknowledge PD fellowships awarded by FAPERJ.

E.S. and F.J.L. extend their thanks to the Paul Scherrer Institute in Switzerland for providing access to µSR experiments, with special recognition to Dr. Hubertus Luetkens for his invaluable assistance.

**Author contributions**

E.S.: the concept of the project, the performance of µSR measurements, analysis, and interpretation of data, writing of the original draft. L.P. and V.T.: synthesis and characterization (magnetization, XRD, heat capacity) of materials, contributions to manuscript editing. C.P.C.M. and M.A.V.H: Mössbauer spectroscopy measurements and contributions to manuscript editing. E.M.B.S.: supervision of Mössbauer spectroscopy, data interpretation, and manuscript editing. F.J.L.: supervision of the project, the performance of µSR measurements, analysis of the Mössbauer spectra, interpretation of experimental data, and manuscript editing.

**Data availability**

The data supporting this study's findings are available from the corresponding author upon reasonable request.

**Declarations**
**Conflict of interest**
There are no conflicts to declare.